\documentclass[prd,nopacs,preprintnumbers,twocolumn,amsmath,nofootinbib,amssymb,floatfix]{revtex4}
\usepackage{graphicx,color,dcolumn,booktabs,bm}
\usepackage{longtable,lscape}
\usepackage{ulem}
\usepackage{txfonts}
\usepackage{overpic}
\usepackage{amssymb}
\usepackage{epstopdf}
\usepackage{indentfirst}
\usepackage{feynmf}   
\usepackage{slashed}  
\usepackage{cases}
\usepackage{color}
\usepackage{float}
\usepackage{multirow}
\usepackage{epsfig,dsfont,amssymb,amsmath,amsfonts,amsbsy,mathrsfs}
\usepackage{comment}

\graphicspath{{Figures/}} %

\usepackage{hyperref}
\hypersetup{colorlinks,citecolor=blue,anchorcolor=red,menucolor=red, linkcolor=red,filecolor=red,runcolor=red,urlcolor=blue}

\makeatletter
\@addtoreset{equation}{section}
\makeatother

\allowdisplaybreaks

\begin{document}

\title{The odd-parity strange baryons $\Sigma\,(\frac{1}{2}^-)$ below 1.8 GeV with Hamiltonian effective field theory}
\author{Zhong-Lin Ma$^{1,2,3}$}
\author{Zhan-Wei Liu$^{1,2,3}$}\email{liuzhanwei@lzu.edu.cn}
\author{Jiong-Jiong Liu$^{4,1}$}
\affiliation{
$^1$School of Physical Science and Technology, Lanzhou University, Lanzhou 730000, China\\
$^2$Research Center for Hadron and CSR Physics, Lanzhou University and Institute of Modern Physics of CAS, Lanzhou 730000, China\\
$^3$Lanzhou Center for Theoretical Physics, MoE Frontiers Science Center for Rare Isotopes, Key Laboratory of Quantum Theory and Applications of MoE, Key Laboratory of Theoretical Physics of Gansu Province, Gansu Provincial Research Center for Basic Disciplines of Quantum Physics, Lanzhou University, Lanzhou 730000, China\\
$^4$State Grid Xinjiang Electric Power Research Institute, Urumqi
830011, Xinjiang, China
}

\begin{abstract}
We examine the spectrum of the $\Sigma\,(\frac{1}{2}^-)$ family based on the experimental $K^-p$ scattering data and lattice QCD simulations within the Hamiltonian Effective Field Theory. Especially, two different scenarios are constructed in order to clarify whether there is one or two $\Sigma\,(\frac{1}{2}^-)$ resonances with masses around 1.5$\sim$1.7 GeV. The relevant lattice QCD data support our scenario with two resonance poles at $1687-110\,i$ and $1714-14\,i$ MeV in which the bare strange triquark core plays an important role. We also show an extra clear cusp structure around 1.4 GeV in our scattering T matrices associated with the odd-parity strange baryons. 
\end{abstract}

\maketitle

\section{introduction}\label{sec1}
The study of the spectroscopy and structure of low-lying resonances has been a pivotal subject in hadron physics for more than half a century. The traditional quark model has achieved considerable success in characterizing the majority of hadronic resonances. However, it encounters significant challenges in explaining certain resonances observed in experiments~\cite{Sumihama:2018moz,LHCb:2015yax,LHCb:2019kea,LHCb:2021auc,LHCb:2021vvq,Feijoo:2021zau,BESIII:2024jgy,Wang:2024kke}. Taking the $J^P=1/2^-$ baryons as examples, the nonstrange $N(1535)$ has a larger mass than the $\Lambda(1405)$ with strangeness contrary to the prediction of naive quark model. This ``mass inversion problem'' can be reasonably solved if treating $\Lambda(1405)$ mainly as a $\bar KN$ molecule.

The study of singly-strange hyperons can provide a unique perspective for exploring the properties of the non-perturbative regime of strong interactions. Within the framework of the unquenched quark model, there may be a $\Sigma\left(\frac{1}{2}^-\right)$ resonance around 1.4 GeV, serving as a partner state to the $\Lambda(1405)$~\cite{Helminen:2000jb,Zhang:2004xt}. The existence of a low energy $\Sigma\left(\frac{1}{2}^-\right)$ resonance was also predicted with the S wave $\bar{K}N$ interactions investigated by using the coupled channel chiral unitary approach ~\cite{Jido:2003cb,Oller:2000fj,Khemchandani:2018amu,Kamiya:2016jqc,Oller:2006jw,Li:2024tvo,Zou:2010tc,Lu:2022hwm,He:2025vij,Lin:2025pyk,Roca:2013cca}.

The $\Sigma(1620)$ and $\Sigma(1750)$ resonances should  also exist in the $\Sigma\left(\frac{1}{2}^-\right)$ family, according to a multi-channel analysis of $\bar{K}N$ interactions~\cite{Kim:1971zxa,Langbein:1972uhb,Jones:1974si}. A recent study~\cite{Guo:2025mha} utilized the effective Lagrangian approach to analyze the $K_Lp \rightarrow \Sigma^0 \pi^+$ process, emphasizing the significance of the $\Sigma(1620)$ in this reaction. An analysis of the $\bar{K}N \rightarrow \Lambda \pi$ reaction using the effective Lagrangian method found no evidence for such a resonance~\cite{Gao:2012zh}. In contrast, studies based on the chiral quark model suggest that the contribution of the $\Sigma(1620)$ cannot be ruled out, particularly with higher-precision data~\cite{Zhong:2013oqa}. The couplings of the $\Sigma(1750)$ resonance to the $\pi\Lambda$ and $\pi\Sigma$ channels were analyzed based on $\text{SU(6)} \otimes \text{O(3)}$ symmetry~\cite{Faiman:1975yz}. The dynamical coupled channel model was employed to establish the energy spectrum of $\Sigma$ hyperon resonances and to extract resonance parameters~\cite{Kamano:2014zba,Kamano:2015hxa}. A recent review of the low-energy $\Sigma\left(\frac{1}{2}^-\right)$ states can be found in Ref.~\cite{Wang:2024jyk}.

Experimentally, most studies on $\Sigma$ hyperons originated from $K^-p$ scattering experiments~\cite{Rutherford-London:1975zvn,Prakhov:2008dc,Alston-Garnjost:1976zzn,Armenteros:1969lza,Armenteros:1970eg}. In the low-lying $\Sigma\left(\frac{1}{2}^-\right)$ spectrum, the Review of Particle Physics (RPP) lists only a one-star $\Sigma(1620)$ and a three-star $\Sigma(1750)$ ~\cite{ParticleDataGroup:2024cfk}. 
The $\Sigma\left(\frac{1}{2}^-\right)$ resonance below 1.5 GeV appeared once in the 2019 version of the RPP with one-star as the $\Sigma(1480)$~\cite{ParticleDataGroup:2018ovx}. However, the Belle Collaboration analyzed the invariant mass distributions of $\pi^+\Lambda$ and $\pi^-\Lambda$ in $\Lambda_c^+ \rightarrow \Lambda \pi^+ \pi^+ \pi^-$ decay process and reported the mass and width of the possible resonance as $1434.3 \pm 0.6 (\text{stat}) \pm 0.9 (\text{syst}) \ \text{MeV}$ and $11.5 \pm 2.5 (\text{stat}) \pm 5.3 (\text{syst}) \ \text{MeV}$, respectively~\cite{Belle:2022ywa}. Based on $e^+e^-$ annihilation data, the BESIII Collaboration analyzed the $\Lambda_c^+ \rightarrow \Lambda \pi^+ \eta$ decay process employing the resonance mass 1380 MeV and width 120 MeV~\cite{BESIII:2024mbf}. Several processes have been proposed as promising channels for searching for $\Sigma\left(\frac{1}{2}^-\right)$  resonances, including $\gamma p \to K^+ \Sigma \pi$~\cite{CLAS:2013rjt}, $\Lambda_c^+ \to \gamma \pi^+ \Lambda$~\cite{Wang:2024ewe}, $\Lambda_c^+ \to \eta \pi^+ \Lambda$~\cite{Lyu:2024qgc,Xie:2017xwx}, $\bar{K} p \to \Lambda \pi^+ \pi^-$~\cite{Wu:2009tu}, $\Xi_c^+ \to \Lambda \bar{K}^0 \pi^+$~\cite{Li:2025exm}, $J/\psi \to \Lambda \bar{\Lambda} \pi$~\cite{Huang:2024oai}, $p\bar{p}\rightarrow \bar{\Lambda }\Sigma \eta $~\cite{Xu:2024xso}, and so on.

The mass spectrum of the low lying $\Sigma\left(\frac{1}{2}^-\right)$ resonances remains unclear, primarily due to limited statistics and significant background uncertainties. There are significant discrepancies in the extracted pole positions among different groups. For instance, the pole for $\Sigma\left(\frac{1}{2}^-\right)$ around 1.7 GeV was reported as 1689-103i MeV in Ref.~\cite{Sarantsev:2019xxm} versus 1708-79i MeV in Ref.~\cite{Zhang:2013sva}. We are not even sure how many low-lying $\Sigma\left(\frac{1}{2}^-\right)$ resonances there should be.

One of the reasons for the still unclear understanding of $\Sigma\left(\frac{1}{2}^-\right)$ family is that their contributions are not dominant in most processes, unlike those of resonances with other quantum numbers. For example, in the $\Lambda_c^+ \rightarrow \Lambda \pi^+ \eta$ decay, the peak of $\Sigma(1385)$ with $J^P = 3/2^+$ dominates rather than that of the $\Sigma\left(\frac{1}{2}^-\right)$ around 1.4 GeV~\cite{BESIII:2024mbf}. Luckily, the $\Sigma\left(\frac{1}{2}^-\right)$ mass spectrum can be simulated with lattice QCD without other quantum number entanglements in some way, which provides us the unique information to investigate the hyperon resonances.

The Hamiltonian Effective Field Theory (HEFT) can analyze the baryon spectra obtained with the lattice QCD in finite volume and extrapolate them to the physical world \cite{Liu:2016wxq,Hall:2014uca,Liu:2023xvy,Liu:2015ktc,Abell:2023nex,Zhuge:2024iuw,Wu:2014vma,Hockley:2024ipz,Han:2025gkp}.  The $\Lambda(1405)$ is interpreted mainly as a $\bar{K}N$ bound state with HEFT~\cite{Liu:2016wxq,Hall:2014uca}, but the triquark core from the naive quark model still exists and plays an important role in the internal structure of $\Lambda(1670)$ in our approach~\cite{Liu:2023xvy}. The finite-volume energy levels of $\Lambda\left(\frac{1}{2}^-\right)$ with HEFT in Ref.~\cite{Liu:2023xvy} are in close agreement with the latest lattice QCD simulations for the isospin $I = 0$ hyperons presented in Refs.~\cite{BaryonScatteringBaSc:2023ori,BaryonScatteringBaSc:2023zvt}.

In this work, we will investigate the structure of odd-parity strange baryons with the isospin $I = 1$ using HEFT. The BGR Collaboration extracted the $\Sigma\left(\frac{1}{2}^-\right)$ mass spectrum with diverse interpolators using two dynamical light chirally improved quarks and a valence strange quark~\cite{Engel:2013ig,Engel:2011aa,Engel:2012qp}, and the simulations span seven ensembles with pion masses ranging from 255 to 596 MeV. In addition to the lattice QCD data, we also consider the $K^-p$ scattering cross-section points up to laboratory momenta of 800 MeV in experiments. The structures of the $\Sigma\left(\frac{1}{2}^-\right)$ baryons are investigated by combining these analyses.

The paper is organized as follows. We provide the HEFT framework for the $\Sigma\left(\frac{1}{2}^-\right)$ family in Sec.~\ref{sec2} and present the numerical results and discussion in Sec.~\ref{sec3}. A short summary is given in Sec.~\ref{SUM}.

\section{Framework}
\label{sec2}
In this section, we briefly present HEFT for the study of $\Sigma$ resonances. One part is for the cross sections of $K^-p$ scatterings, and the other is for the finite volume spectra. 

In the center-of-mass frame, the Hamiltonian for the single-strange baryon system with the isospin $I$ is
\begin{equation}
\label{Hamiltonian}
H^{I}=H^{I}_0+H^{I}_{int} \, ,
\end{equation}
Here, $H^{I}_0$ represents the Hamiltonian of the non-interacting part,
\begin{eqnarray}
\label{H0Hamiltonian}
H^{I}_0&=&\sum_{B_0}|B_0\rangle\, m_B\,\langle B_0| \nonumber\\
&&+\sum_{\alpha}\int d^3 k\left|\alpha\left(k\right)\right\rangle\left[\omega_{\alpha_{M}}\left(k\right)+\omega_{\alpha_B}\left(k\right)\right]\left\langle\alpha\left(k\right)\right|,\quad 
\end{eqnarray}
where $m_B$ is the mass of the bare state $B_0$. $\alpha$ refers to the meson-baryon channel, and we consider $\pi\Lambda$, $\pi\Sigma$, $\bar{K}N$, $\eta\Sigma$, and $K\Xi$ for the $I=1$ case. The non-interacting energy of meson (baryon) is written as 
\begin{eqnarray}
\label{freeenergy}
\omega_{\alpha_{M\left(B\right)}}\left(k\right)=\sqrt{m^2_{\alpha_{M\left(B\right)}}+k^2}.
\end{eqnarray}
The interaction Hamiltonian $H^{I}_{int}$ can be decomposed into two components
\begin{equation}
\label{Hii}
H^{I}_{int}=g^{I}+v^{I} \, .
\end{equation}
$g^{I}$ denotes the interaction between a bare states and two-particle basis state
\begin{eqnarray}
\label{bareHamiltonian}
g^{I}=\sum_{B_0,\alpha}\int d^3 k\left\{\left|B_0\right\rangle G^{I\dagger}_{B_0,\alpha}\left(k\right)\left\langle \alpha\left(k\right)\right|+h.c.\right\} \, ,
\end{eqnarray}
and we take the form
\begin{eqnarray}
\label{bareintaction}
G^{I}_{B_0,\alpha}(k)=\frac{\sqrt{3}\,g^I_{B_0,\alpha}}{2\pi f}\sqrt{\omega_{\alpha_M}(k)}\,u(k) \, ,
\end{eqnarray}
where $g^I_{B_0,\alpha}$ is the coupling strength in the isospin $I$ channel, and we take the dipole regulator $u(k,\Lambda)=(1+k^2/\Lambda^2)^{-2}$ with $\Lambda=1$ GeV. The term $v^{I}$ represents the interaction between two meson-baryon states
\begin{eqnarray}
v^{I}=\sum_{\alpha,\beta}\int d^3 kd^3 k^\prime\left|\alpha(k)\right\rangle V^{I}_{\alpha,\beta}\left(k,k^\prime\right)\left\langle\beta(k^\prime)\right| \, ,
\end{eqnarray}
and we use the Weinberg-Tomozawa term with the couplings $g^I_{\alpha,\beta}$~\cite{Weinberg:1966kf,Thomas:1981ps}
\begin{eqnarray}
V^{I}_{\alpha,\beta}\left(k,k^\prime\right)=g^I_{\alpha,\beta}\frac{\left[\omega_{\alpha_M}\left(k\right)+\omega_{\beta_M}\left(k^\prime\right)\right]u\left(k\right)u\left(k^\prime\right)}{8\pi^2f^2\sqrt{2\omega_{\alpha_M}\left(k\right)}
	\sqrt{2\omega_{\beta_M}\left(k^\prime\right)}} \, .
\label{interactionp}
\end{eqnarray}
%


By solving the three-dimensional reduced form of the Bethe-Salpeter equation with the above interactions, we can obtain the T-matrix
\begin{eqnarray}
&&\hspace{-2em}T^{I}_{\alpha,\beta}(k,k^\prime;E)=\widetilde{V}^{I}_{\alpha,\beta}(k,k^\prime;E)\nonumber\\
&&\qquad\quad  +\sum_\gamma
\int q^2dq\frac{\widetilde{V}^{I}_{\alpha,\gamma}(k,q;E)\, T^{I}_{\gamma,\beta}(q,k^\prime;E)}{E-\omega_{\gamma_1}(q)-\omega_{\gamma_2}(q)+i\epsilon},
\end{eqnarray}
where
\begin{eqnarray}
\widetilde{V}^{I}_{\alpha,\beta}(k,k^\prime ;E)&=&\sum_{B_0}\frac{G^{I\dagger}_{B_0,\alpha}(k)\,G^{I}_{B_0,\beta}(k^\prime)}
{E-m_B^0}+V^I_{\alpha,\beta}(k,k^\prime) \, .\qquad 
\end{eqnarray}
One should note that the $K^-p$ scattering T-matrix $T_{\alpha,\beta}$ should be written as the linear combination of $T^{I}_{\alpha,\beta}$, that is, $T_{\alpha,\beta}(k,k^\prime;E)=a\, T^{I=0}_{\alpha,\beta}(k,k^\prime;E)+b\, T^{I=1}_{\alpha,\beta}(k,k^\prime;E)$, where $a$ and $b$ are the corresponding Clebsch-Gordan coefficients. Using the T-matrix one can easily extract the cross section
\begin{eqnarray}
\sigma_{\alpha,\beta}&=&\frac{4\pi^3 k_{\alpha\,{\rm cm}}\,\omega_{\alpha_M\,{\rm cm}}\,\omega_{\alpha_B\,{\rm cm}}\,\omega_{\beta_M\,{\rm cm}}\,\omega_{\beta_B\,{\rm cm}}}{E^2_{\rm cm}\,k_{\beta\,{\rm cm}}}\,\nonumber\\
&&\qquad \times |T_{\alpha,\beta}(k_{\alpha\,{\rm cm}},k_{\beta\,\rm cm};E_{\rm cm})|^2 \, , 
\label{crosssection}
\end{eqnarray}
where the subscript ``${\rm cm}$'' refers to the center-of-mass momentum frame. Furthermore, the resonance poles can be determined by searching for those of the T-matrix in the complex plane.


Next, we list our finite volume formalism for comparing with lattice QCD results. In a box of length $L$,  the particle momenta are discretized as $k_n=2\pi\sqrt{n}/L$, with $n=n_x^2+n_y^2+n_z^2$, where $n=0,1,2,...$. In the finite volume, the free Hamiltonian for the isospin-1 system is discretized as 

\begin{eqnarray}
\mathcal{H}_{0}^{1}&=&\text{diag} \left\{
m_{\Sigma_0},
\omega_{\pi\Lambda}\left(k_0\right),
\omega_{\pi\Sigma}\left(k_0\right),
\omega_{\bar{K}N}\left(k_0\right),
\omega_{\eta\Sigma}\left(k_0\right),\right. \nonumber \\
& & \left. 
\omega_{K\Xi}\left(k_0\right),  
\omega_{\pi\Lambda}\left(k_1\right),
\omega_{\pi\Sigma}\left(k_1\right),\
\ldots
\right\}.
\end{eqnarray}
%
and the interacting part is given by $\mathcal{H}_{int}^1 =$
\begin{eqnarray}
 \left(
\begin{array}{ccccc}
0 & \mathcal{G}^1_{\Sigma_0,\pi\Lambda}\left(k_0\right) &\mathcal{G}^1_{\Sigma_0,\pi\Sigma}\left(k_0\right) &   \cdots \\
\mathcal{G}^1_{\Sigma_0,\pi\Lambda}\left(k_0\right) &
 \mathcal{V}^1_{\pi\Lambda,\pi\Lambda}\left(k_0,k_0\right) & \mathcal{V}^1_{\pi\Lambda,\pi\Sigma}\left(k_0,k_0\right) &  \cdots \\
 \mathcal{G}^1_{\Sigma_0,\pi\Sigma}\left(k_0\right) &
 \mathcal{V}^1_{\pi\Sigma,\pi\Lambda}\left(k_0,k_0\right) & \mathcal{V}^1_{\pi\Sigma,\pi\Sigma}\left(k_0,k_0\right) &  \cdots \\
 \mathcal{G}^1_{\Sigma_0,\bar{K}N}\left(k_0\right) &
 \mathcal{V}^1_{\bar{K}N,\pi\Lambda}\left(k_0,k_0\right) & \mathcal{V}^1_{\bar{K}N,\pi\Sigma}\left(k_0,k_0\right) &  \cdots \\
\mathcal{G}^1_{\Sigma_0,\eta\Sigma}\left(k_0\right) &
 \mathcal{V}^1_{\eta\Sigma,\pi\Lambda}\left(k_0,k_0\right) & \mathcal{V}^1_{\eta\Sigma,\pi\Sigma}\left(k_0,k_0\right) &  \cdots \\
 \mathcal{G}^1_{\Sigma_0,K\Xi}\left(k_0\right) &
 \mathcal{V}^1_{K\Xi,\pi\Lambda}\left(k_0,k_0\right) & \mathcal{V}^1_{K\Xi,\pi\Sigma}\left(k_0,k_0\right) &  \cdots \\
 \mathcal{G}^1_{\Sigma_0,\pi\Lambda}\left(k_1\right) &
 \mathcal{V}^1_{\pi\Lambda,\pi\Lambda}\left(k_1,k_0\right) & \mathcal{V}^1_{\pi\Lambda,\pi\Sigma}\left(k_1,k_0\right) &  \cdots \\
\vdots &
  \vdots &
\vdots& \ddots  \\
\end{array}
\right),\quad
\end{eqnarray}
where
\begin{eqnarray}
\mathcal{G}^1_{\Sigma_0,\alpha}&=&\sqrt{\frac{C_3\left(n\right)}{4\pi}}\left(\frac{2\pi}{L}\right)^{3/2}G_{\Sigma_0,\alpha}^1\left(k_n\right),\\
\mathcal{V}^1_{\alpha,\beta}\left(k_n,k_m\right)&=&\frac{\sqrt{C_3\left(n\right)C_{3}\left(m\right)}}{4\pi}\left(\frac{2\pi}{L}\right)^3V^1_{\alpha,\beta}\left(k_n,k_m\right) \, .\qquad
\end{eqnarray}
Here, $C_{3}(n)$ denotes the number of ways one can sum the squares of three integers to equal $n$. By solving the eigenvalue equation of Hamiltonian $\mathcal{H}=\mathcal{H}_0+\mathcal{H}_I$, the eigenvalues and eigenvectors can be used to analyze the lattice QCD results. For more details of HEFT, we refer to the study of the $\Lambda\left(\frac{1}{2}^-\right)$ resonances ~\cite{Liu:2023xvy}.

\section{Numerical Results AND DISCUSSION}\label{sec3}
In this section, we investigate the $\Sigma\left(\frac{1}{2}^-\right)$ baryons below 1.8 GeV based on analyzing the $K^-p$ scattering data and the lattice QCD spectra. We consider two scenarios: one is that the bare triquark core $\Sigma_0$ does not play a role in this energy region and these $\Sigma\left(\frac{1}{2}^-\right)$ resonances are dynamically generated by the interaction among the channels $\pi\Lambda$, $\pi\Sigma$, $\bar{K}N$, $\eta\Sigma$, and $K\Xi$, and the other is that the bare triquark core from the quark model is additionally introduced. We initially determine the parameters of the Hamiltonians by fitting the $K^-p$ scattering cross section data. Once these parameters are obtained, we are enabled to compute the finite volume energy spectra and compare with the lattice QCD data.

\subsection{Cross section}

We constrain the parameters by fitting the scattering cross-section data for the $K^-p\rightarrow K^-p$, $K^-p\rightarrow \bar{K}^0n$, $K^-p\rightarrow \pi^0\Lambda^0$, $K^-p\rightarrow \pi^-\Sigma^+$ and $K^-p\rightarrow \pi^+\Sigma^-$ in the infinite volume. These experimental data have been analyzed in our previous work \cite{Liu:2023xvy} where the $I=0$ $\Lambda$ resonances were focused on. Since the $\Lambda\left(\frac{1}{2}^-\right)$ lattice QCD spectra were also fitted well\cite{Liu:2023xvy}, in this work we keep the parameters same as in Ref. \cite{Liu:2023xvy} for the $I=0$ sector.

\begin{figure*}[tbp]
\flushleft
\includegraphics[width=3.7cm,height=3.45cm]{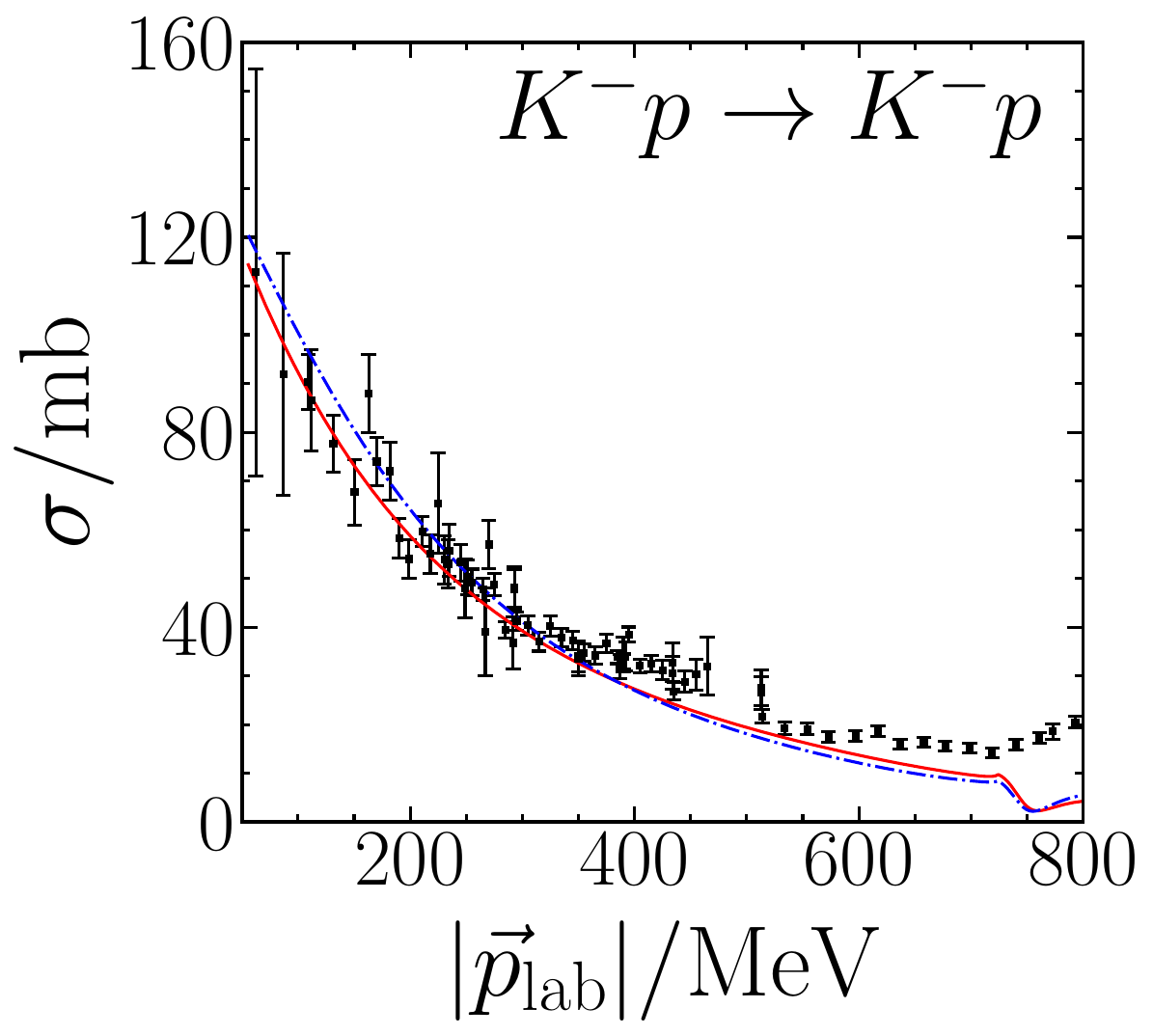} 
\includegraphics[width=3.4cm]{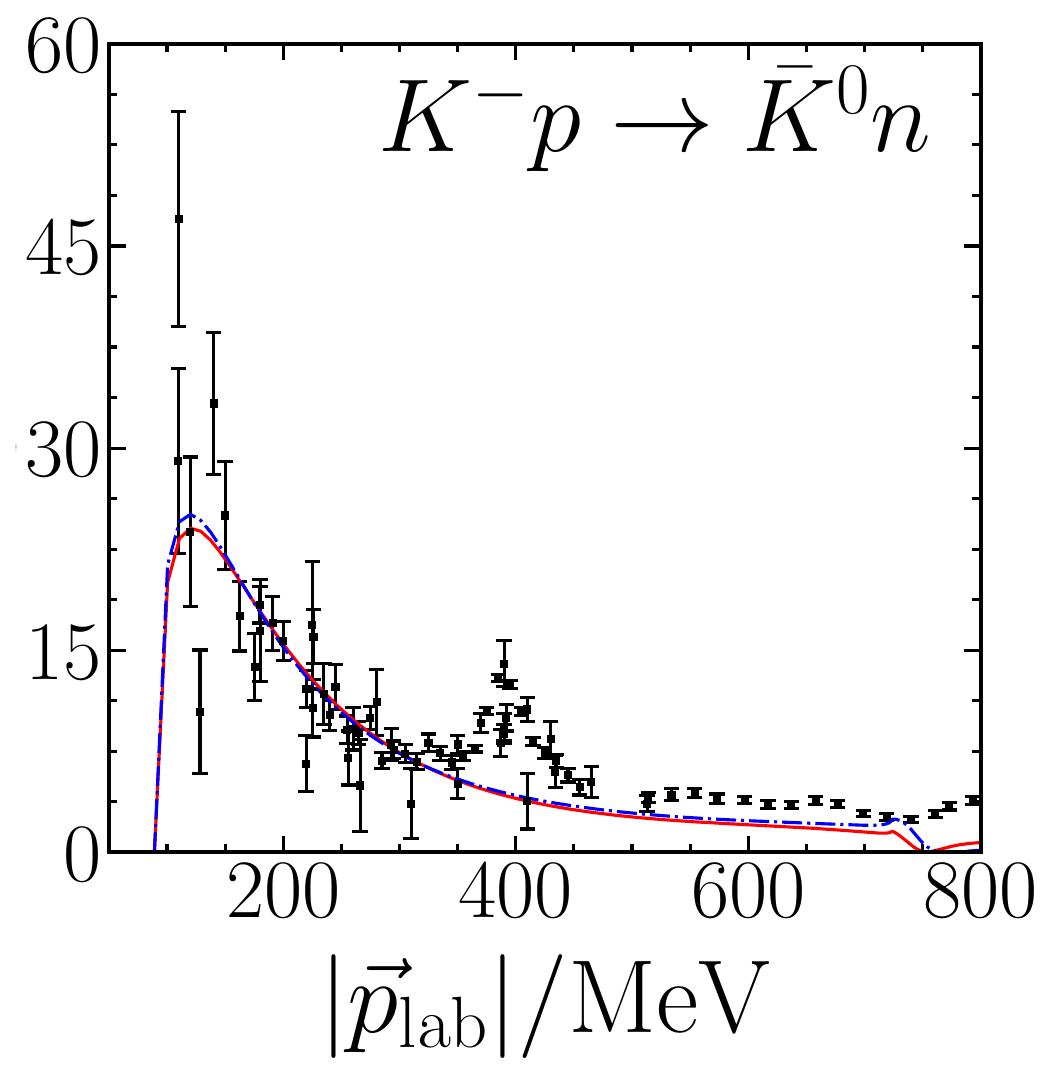} 
\includegraphics[width=3.4cm]{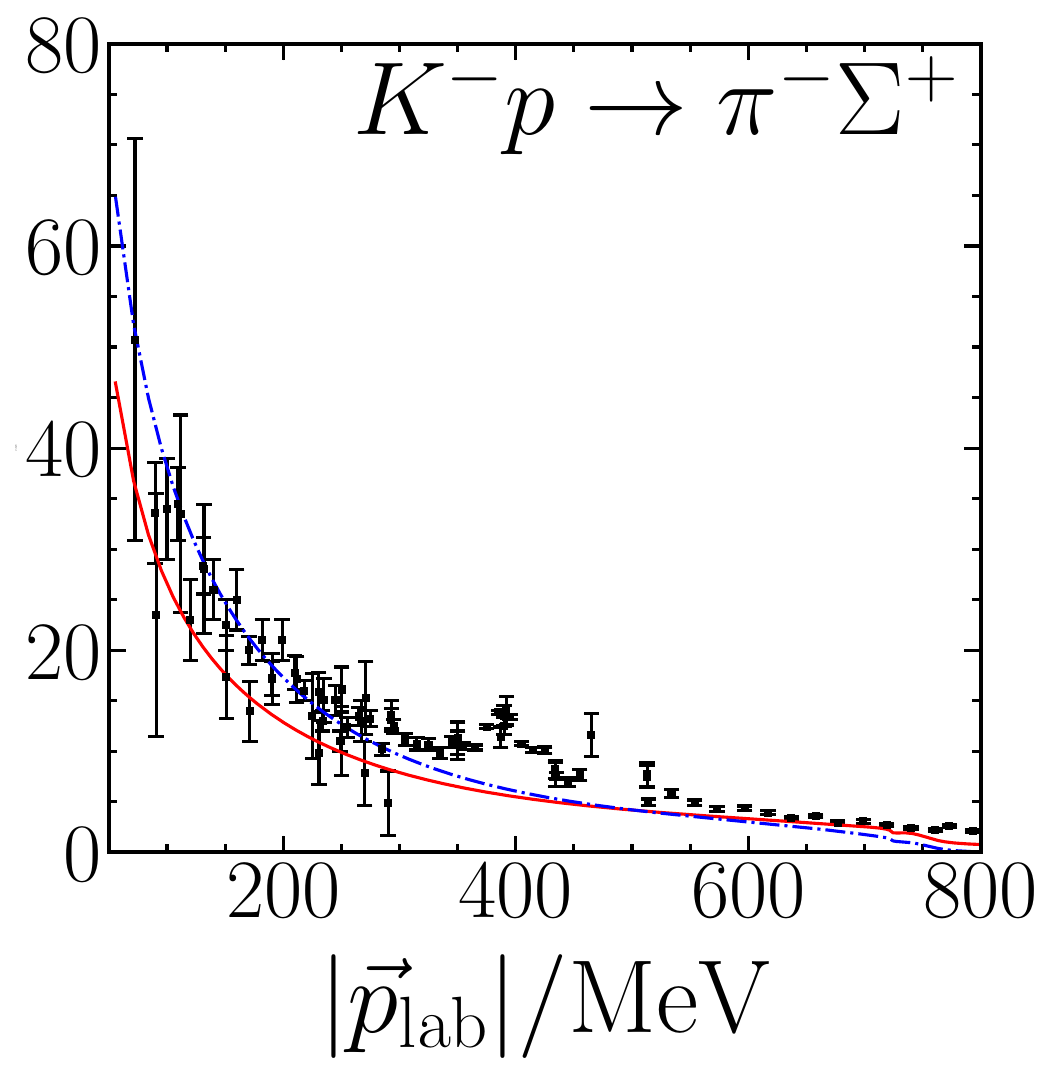} 
\includegraphics[width=3.4cm]{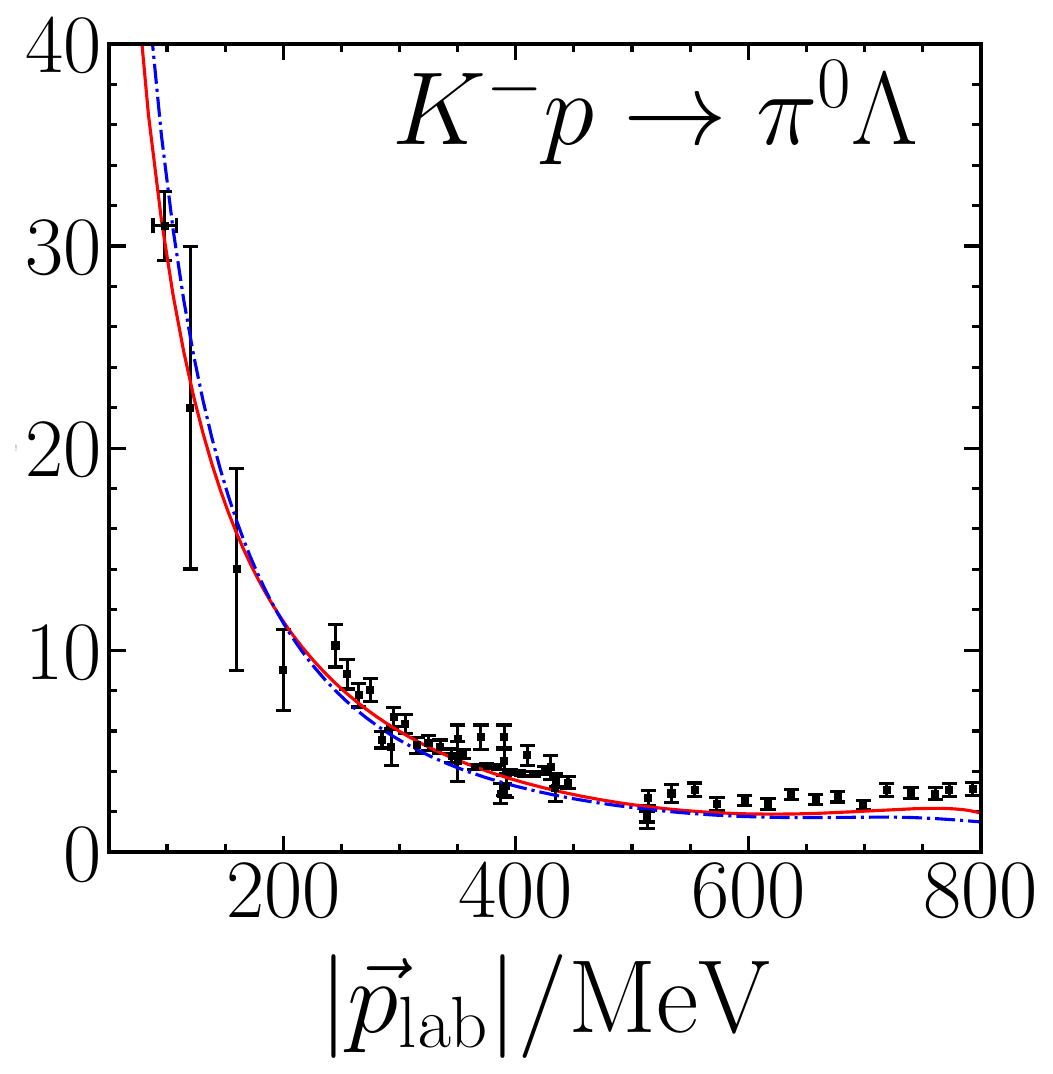}
\includegraphics[width=3.4cm,height=3.45cm]{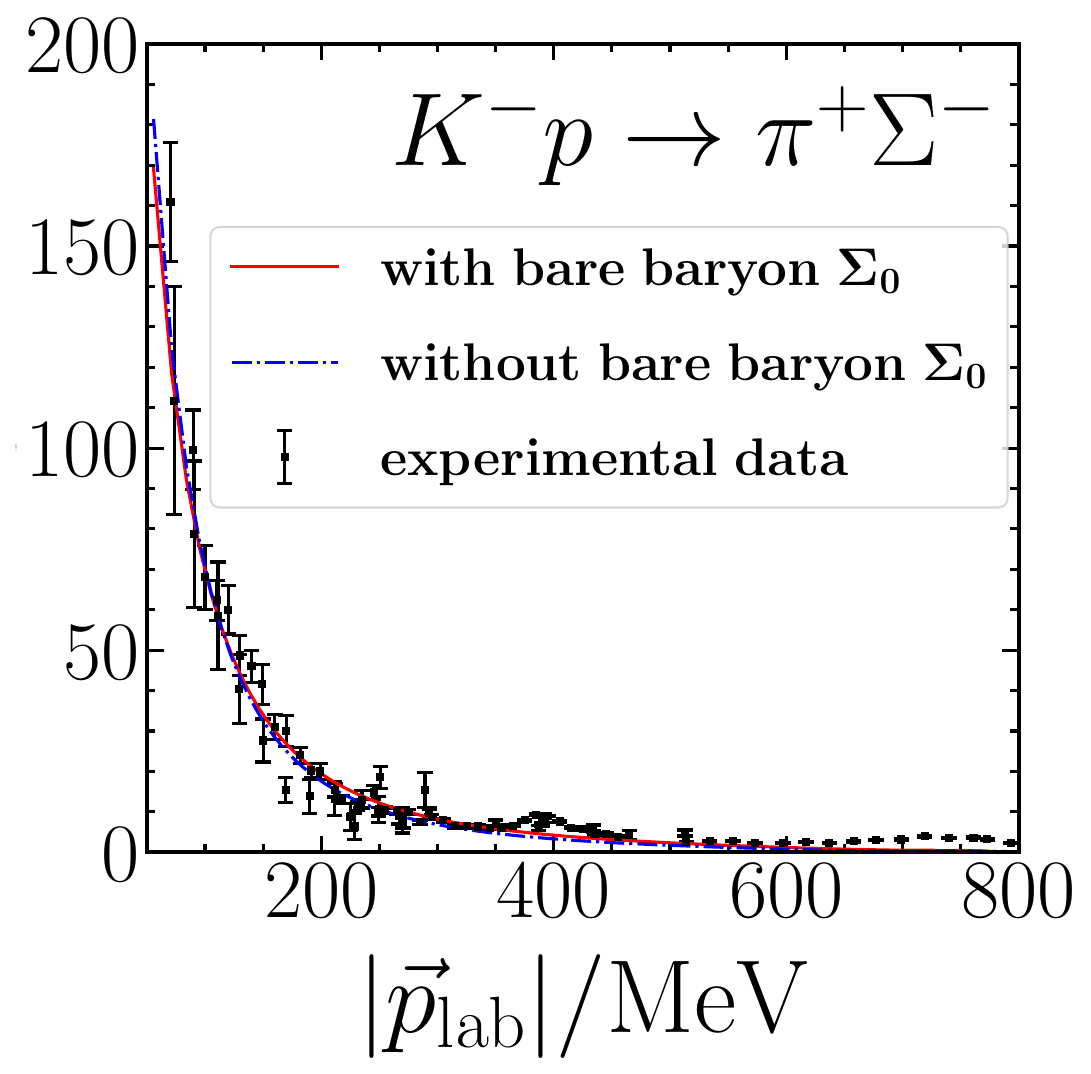}

\caption{The experimental data and fitted results for the $K^-p$ scattering cross sections.  The red solid and the blue dot-dashed lines represent our results with and without the bare baryon component $\Sigma_0$, respectively. The experimental data points are taken from Refs.~\cite{Piscicchia:2022wmd,Abrams:1965zz,Sakitt:1965kh,Kim:1965zzd,Mast:1975pv,Bangerter:1980px,Ciborowski:1982et,Evans:1983hz,Mast:1974sx,Nordin:1961zz,Berley:1996zh,	FerroLuzzi:1962zza,Watson:1963zz,Eberhard:1959zz}. The peaks near 400 MeV are associated with the $I=0$ D-wave $\Lambda(1520)$ resonance which is not included in this work as discussed in the text.}\label{fitting}
		\end{figure*}

The bare baryon $\Lambda_0$ with $I=0$ was considered in Ref. \cite{Liu:2023xvy} while the bare $\Sigma_0$ with $I=1$ was not, and the cross sections therein are plotted in Fig. \ref{fitting} as blue dot-dashed lines. For the other scenario, we introduce both the bare $\Lambda_0$ and $\Sigma_0$, and the refitted $K^-p$ cross sections are shown in Fig. \ref{fitting} as red solid lines. The parameters for the $I=1$ case are summarized in Table \ref{coupling}.

This study focuses on baryons with $J^P=1/2^-$ and the strangeness $S = -1$, and the interactions with other quantum numbers are not considered currently.  We note a discrepancy between our fitted data and the experimental data for $K^-p\rightarrow \bar{K}^0n$ and $K^-p\rightarrow \pi^-\Sigma^+$ near $p_{lab} = 400 \, $MeV. As discussed in Refs.~\cite{Zhong:2013oqa,Lutz:2001yb,He:1993et,Zhong:2008km}, such discrepancies can be eliminated by accounting for P wave and D wave contributions or hyperons like $\Lambda(1520)$ which would not couple with $1/2^-$ resonances.

As can be observed from Fig.~\ref{fitting}, both scenarios provide similarly good descriptions of the scattering reaction data, making it impossible to determine which hypotheses better characterizes the resonance structure solely based on the cross sections. Further analysis is required to determine whether the structure of $\Sigma\left(\frac{1}{2}^-\right)$ includes a bare state component.

		\renewcommand\tabcolsep{0.3cm}
		\renewcommand{\arraystretch}{1.8}
		\begin{table}[!tbp]
\caption{The bare mass and couplings constrained by the $K^-p$ cross sections for the two scenarios. Error estimation for the scenario with bare baryon $\Sigma_0$, obtained by varying the regulator parameter $\Lambda$ within its allowed range $1.0^{+0.1}_{-0.1}$ GeV as discussed in Sec.~\ref{jia}.
}
\centering
\label{coupling}
\begin{tabular}{cccccccccccc}
	\toprule[0.3pt]
	\hline
	\hline
	& Without bare baryon $\Sigma_0$ \cite{Liu:2023xvy}&With bare baryon $\Sigma_0$\\
	
\hline
$m_{\Sigma_0}$ &  & $1720^{+10}_{-9} $ MeV\\
$g_{\Sigma_0, \bar{K} N}^{1}$ & - & $0.050^{+0.007}_{-0.014} $ \\
$g_{\Sigma_0, \pi\Sigma}^{1}$ & - & $0.010^{+0.002}_{+0.010} $ \\
$g_{\Sigma_0, \pi\Lambda}^{1}$ & - & $-0.070^{-0.001}_{+0.029} $ \\
$g_{\Sigma_0, \eta\Sigma}^{1}$ & - & $-0.080^{+0.008}_{+0.008} $ \\
$g_{\Sigma_0, K\Xi}^{1}$ & - & $0.060^{-0.009}_{+0.031}$ \\
$g_{\bar{K} N, \bar{K} N}^{1}$ & -0.001 & $2.200^{+0.050}_{+0.470}$\\
$g_{\bar{K} N, \pi\Sigma}^{1}$ & 0.985 & $1.660^{-0.165}_{+0.191} $ \\
$g_{\bar{K} N, \pi\Lambda}^{1}$ & 0.990 & $-1.300^{+0.119}_{-0.400} $  \\
$g_{\bar{K} N, \eta\Sigma}^{1}$ & 1.500 & $-3.000^{+0.100}_{-0.924} $ \\
$g_{\pi\Sigma, \pi\Sigma}^{1}$ & -0.001 & $2.800^{+0.004}_{+0.001} $ \\
$g_{\pi\Sigma, K\Xi}^{1}$ & -1.341 & $-2.900^{+0.055}_{-0.500} $ \\
$g_{\pi\Lambda, K\Xi}^{1}$ & 0.011 & $-0.250^{-0.005}_{-0.330} $ \\
$g_{\eta\Sigma, K\Xi}^{1}$ & 0.001 & $-0.140^{-0.085}_{-0.100} $ \\
$g_{K\Xi, K\Xi}^{1}$ & -3.700 & $-2.550^{+0.766}_{-0.500} $ \\

	\hline
	\hline
	\bottomrule[0.3pt]
\end{tabular}
\end{table}
%

\subsection{Finite-volume spectra of $\Sigma\left(\frac{1}{2}^-\right)$}
\label{finite-volume results}
Our analysis of scattering cross sections reveals that infinite volume results are insufficient to fully reveal the internal structure of $\Sigma\left(\frac{1}{2}^-\right)$. The finite volume analyses will offer more comprehensive insights to better understand the $\Sigma\left(\frac{1}{2}^-\right)$ resonance spectrum. In this subsection, we analyze the lattice QCD data of the $\Sigma\left(\frac{1}{2}^-\right)$ resonance from various ensembles.

Since the lattice QCD results were provided at unphysical quark masses, we need the hadron masses at larger $m_\pi$ and $m_K$. For the masses of ground baryons $N$, $\Lambda$, $\Sigma$, $\Xi$, we take
\begin{equation}\label{eqmBc}
m_{B} = m_{B}|_{\text{phys}} 
    + \alpha_{B} (m_{\pi}^2 - m_{\pi}^2|_{\text{phys}}) 
    + \beta_{B} (m_{K}^2 - m_{K}^2|_{\text{phys}}).
\end{equation}
The mass of the $\eta$ meson is obtained via the Gell-Mann-Okubo mass relation
\begin{equation}
m_\eta=\sqrt{m_\eta^2|_{\text{phys}}+\frac{4}{3}(m_{K}^2-m_{K}^2|_{\text{phys}})-\frac{1}{3}(m_\pi^2-m_\pi^2|_{\text{phys}})}.
\end{equation}

In this work we analyze the lattice QCD spectrum of $\Sigma\left(\frac{1}{2}^-\right)$ in Ref.~\cite{Engel:2013ig} where other baryon spectra were also provided. With the same ensembles the BGR Collaboration showed the linear dependence of $m_K^2$ on $m_\pi^2$~\cite{Engel:2011aa}
\begin{equation}
    m_K^2 = m_K^2|_{\text{phys}} + b_K \left( m_\pi^2 - m_\pi^2|_{\text{phys.}} \right).
\end{equation} 
Therefore, we can use the simplified formula of Eq. (\ref{eqmBc}) to fit the lattice QCD data 
\begin{equation}
m_{B} = m_{B}|_{\text{phys}} 
    + \tilde\alpha_{B} (m_{\pi}^2 - m_{\pi}^2|_{\text{phys}}),
\end{equation}
where $\tilde\alpha_{B}=\alpha_{B}+\beta_{B}b_K$. For the bare state, we introduce 
\begin{equation}
m_{\Sigma_0} = m_{\Sigma_0}|_{\text{phys}} 
    + \tilde\alpha_{\Sigma_0} (m_{\pi}^2 - m_{\pi}^2|_{\text{phys}}).
\end{equation}
We approximately employ the fitted $\tilde\alpha_{\Sigma}$ of ground state $\Sigma$ baryon to constrain the slope of the bare $\Sigma_0$, $ \tilde\alpha_{\Sigma_0} =\tilde\alpha_{\Sigma}= 1.3\,\text{GeV}^{-1}$.

\begin{figure}[tbp]
\center
\includegraphics[width=3.4in]{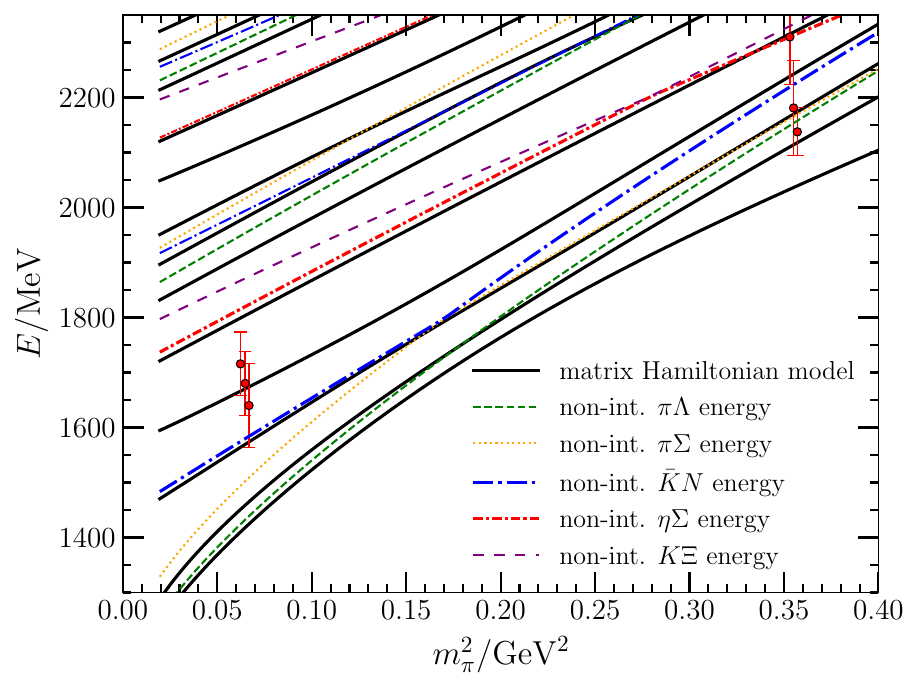}\\
(a) Without bare baryon $\Sigma_0$\\
\vspace{2em}
\includegraphics[width=3.4in]{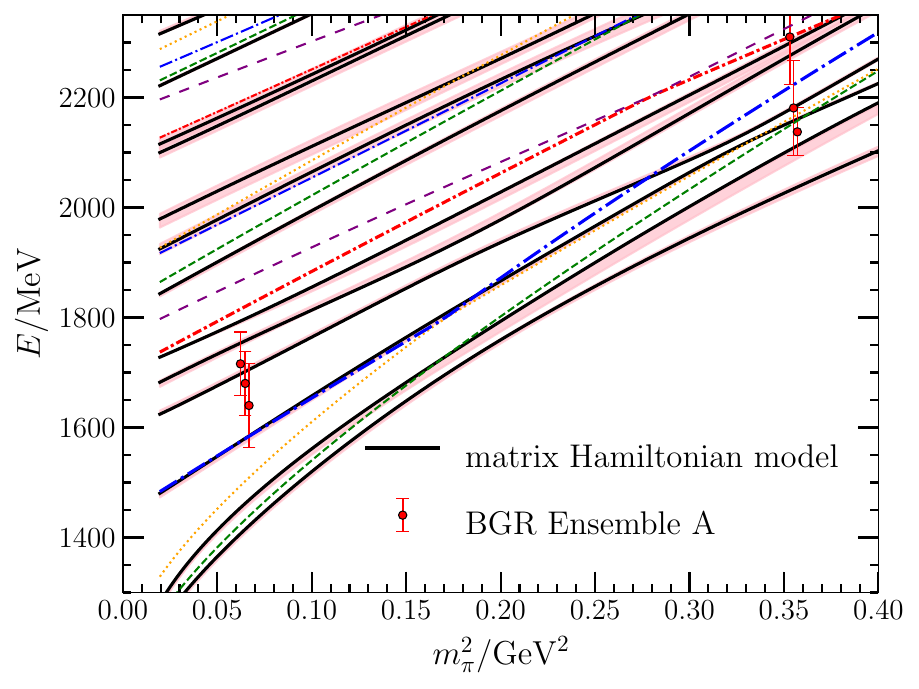}\\
(b) With bare baryon $\Sigma_0$\\
\caption{Finite-volume energy spectra in two scenarios compared to the lattice QCD data with Ensemble A ($L = 2.118\,\text{fm} $) from the BGR group \cite{Engel:2013ig}. The broken lines denote noninteracting meson-baryon energies, while the solid lines denote the eigenenergies obtained from the finite-volume Hamiltonian matrix. The pink shaded area illustrates the uncertainty of the HEFT results obtained by varying the Hamiltonian parameters within the allowed range as described in Sec.~\ref{jia}. For clarity, we shift the higher (lower) lattice QCD datum to the left (right) a little while keeping middle datum unchanged on each pion mass.}

\label{withoutbaresp}
\end{figure}

We are now able to compute the finite volume energy spectra and compare them with lattice QCD results across a range of lattice spacings from 0.1324 fm  to 0.1398 fm based on the parameters in Table~\ref{coupling}. In Fig.~\ref{withoutbaresp}, we illustrate the energy levels of HEFT with the variation of pion mass. The lattice QCD data from Ensemble A in Ref.~\cite{Engel:2013ig} are shown as the red data points with error bars in Fig.~\ref{withoutbaresp}. The HEFT results with the two hypotheses are both presented. We also plot the non-interacting energies of two-body channels as the dashed lines. 

From Fig.~\ref{withoutbaresp}, there are three excited states with the masses around 1.5$\sim$1.7 GeV observed at the smallest pion mass on the lattice, and moreover, two of them are far away from any two-body channel threshold over 1$\sigma$ at least. However, there is only one HEFT energy level between the blue long-dashed-dotted and red short-dashed-dotted lines and not close to them in  Fig.~\ref{withoutbaresp} (a), which indicates that the lattice QCD spectrum of $\Sigma\left(\frac{1}{2}^-\right)$ cannot be interpreted very well in the absence of a bare baryon $\Sigma_0$. As can be seen from Fig.~\ref{withoutbaresp} (b), two HEFT eigenstates have the masses around 1.6$\sim$1.7 GeV at the small pion masses, which is consistent with the lattice QCD data with the inclusion of $\Sigma_0$. Although the HEFT spectra are also different for the large pion masses, the current lattice QCD data from the BGR Collaboration cannot tell which scenario is better. Therefore, the current lattice QCD data at the small pion mass prefer the scenario with the bare baryon $\Sigma_0$ from Fig.~\ref{withoutbaresp}. We will focus on this scenario in the following analysis.

\begin{figure}[tb]
\center

	\includegraphics[width=3.9cm]{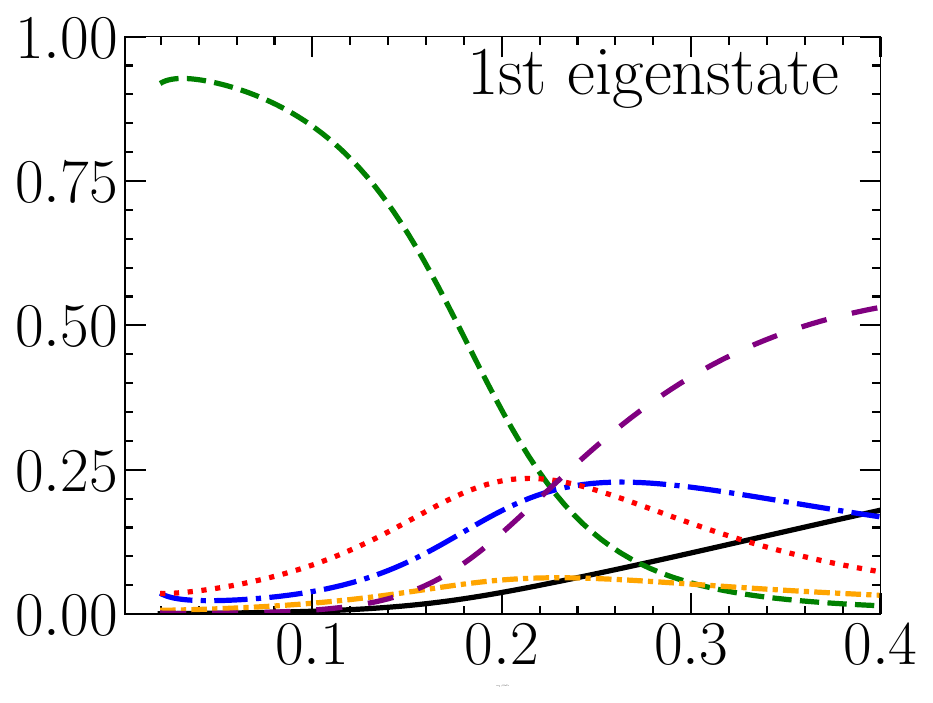}
	\includegraphics[width=3.9cm]{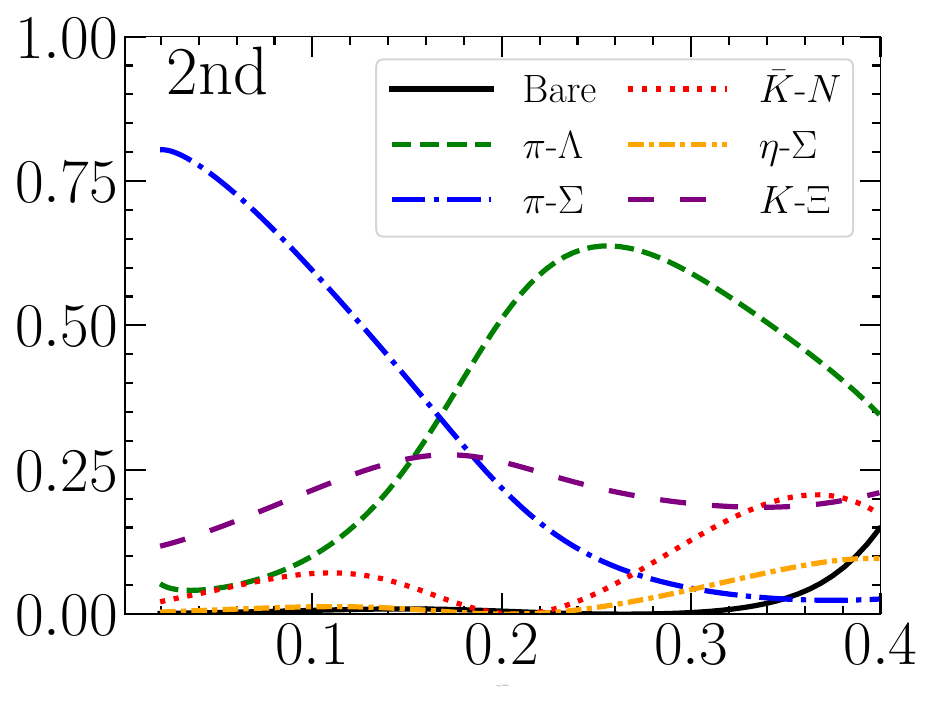}
	\includegraphics[width=3.9cm]{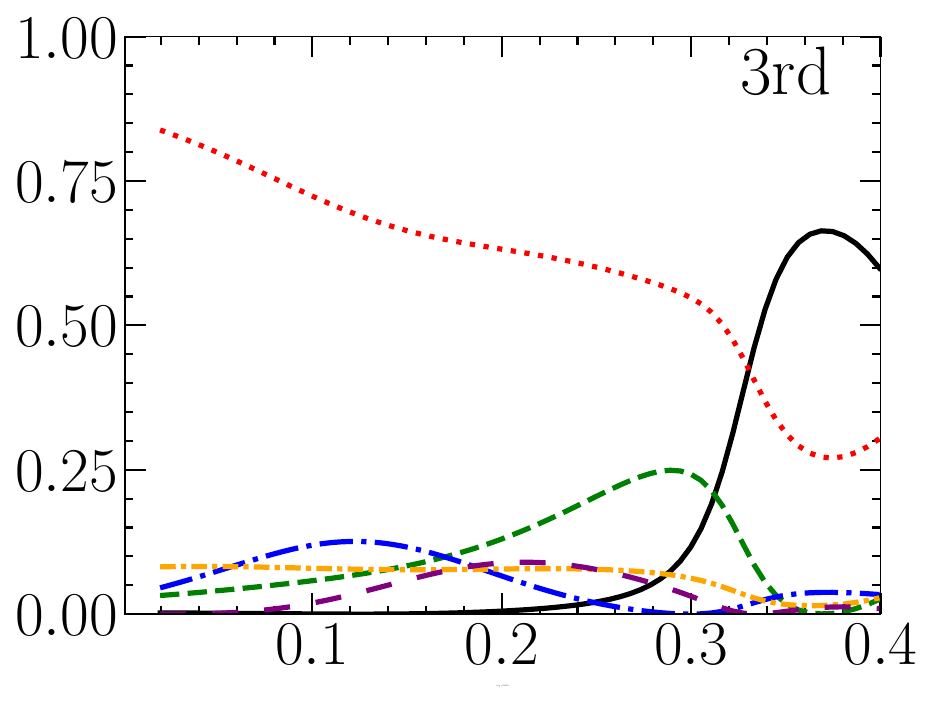}
	\includegraphics[width=3.9cm]{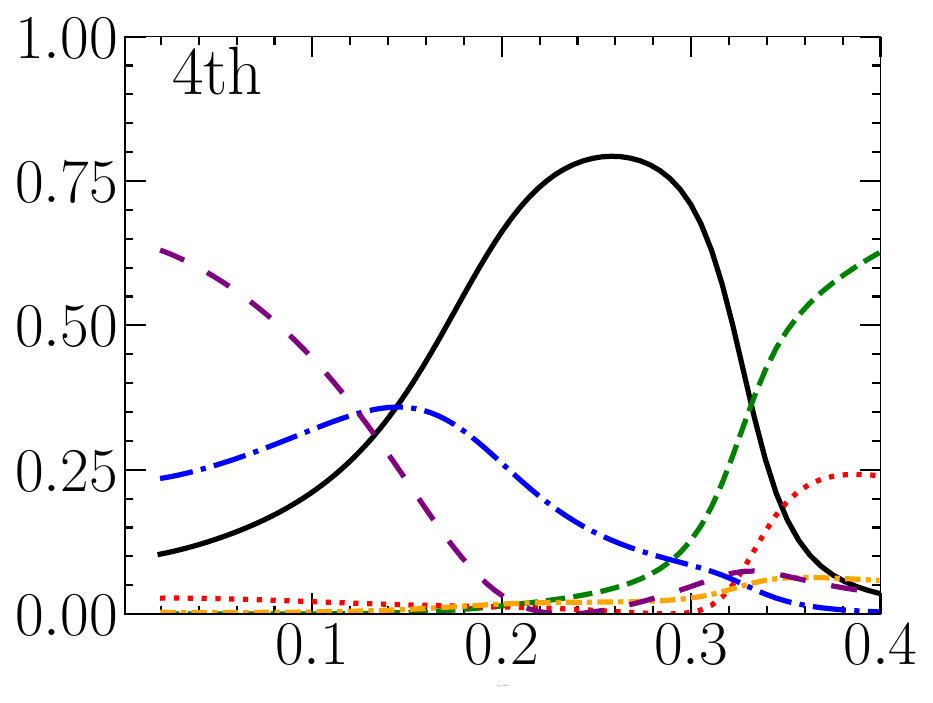}
	\includegraphics[width=3.9cm]{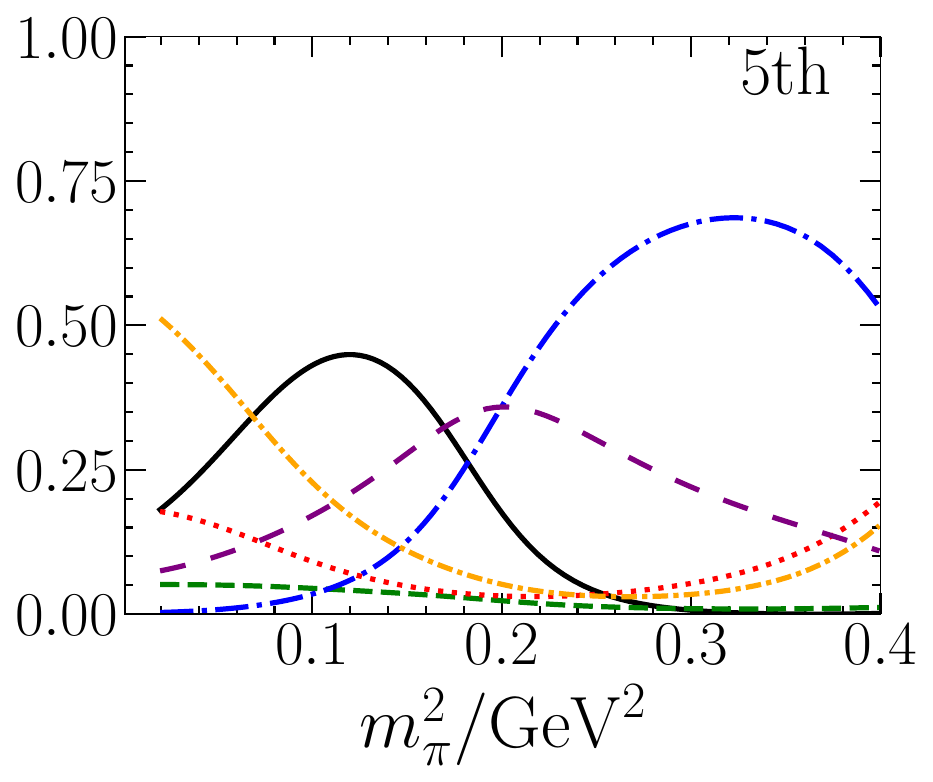}
	\includegraphics[width=3.9cm]{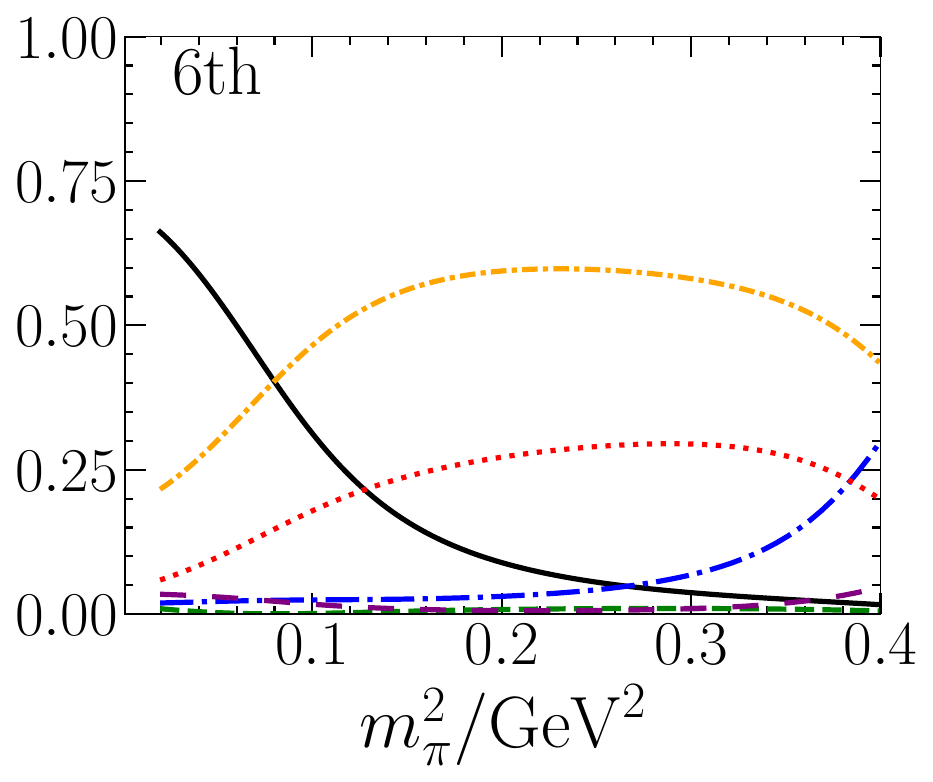}\\
       
\caption{The dependence of the Hamiltonian eigenvector components on the pion mass for the first six eigenstates  under the hypothesis with a bare baryon $\Sigma_0$ .}
\label{vectors}
\end{figure}

In Fig.~\ref{vectors}, we show the composition of the low-lying HEFT eigenstates under the hypothesis including the bare $\Sigma_0$ baryon. The first eigenstate is predominantly composed of $\pi \Lambda$ at small pion mass and becomes dominated by $K \Xi$ as the pion mass increases. The second and third states are mainly $\pi\Sigma$ and $\bar K N$ for the small pion masses, respectively. At the physical pion mass, the fourth energy level is about one hundred MeV below the $K \Xi$ threshold from 
Fig.~\ref{withoutbaresp} (b), but this state is dominated by $K \Xi$, which states that the $K \Xi$ attraction is very strong in our model. The fifth and sixth states are mixed with the bare $\Sigma_0$, $\eta\Sigma$, and so on when $m_\pi$ is not large.

\begin{figure}[tb]
	\includegraphics[width=3.4in]{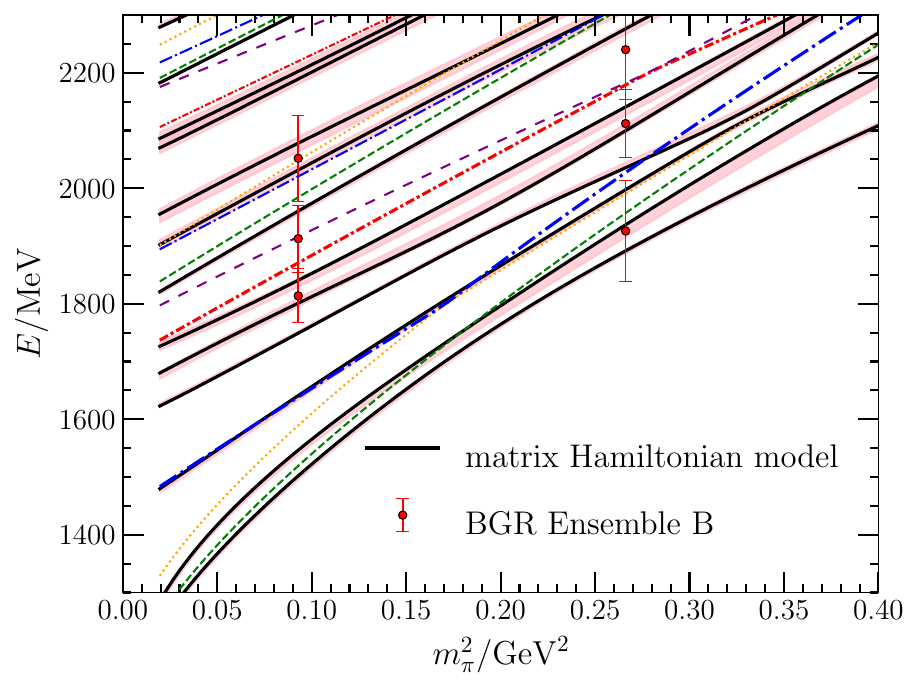}\\
\caption{Finite-volume energy spectrum with the bare baryon $\Sigma_0$ compared to the lattice QCD data with Ensemble B ($L = 2.186\,\text{fm} $) from the BGR group \cite{Engel:2013ig}. The pink shaded area illustrates the uncertainty of the HEFT results obtained by varying the Hamiltonian parameters within the allowed range as described in Sec.~\ref{jia}. See Fig.~\ref{withoutbaresp} for other captions.}
\label{bben}
\end{figure}

\begin{figure}[htb]

	\includegraphics[width=3.4in]{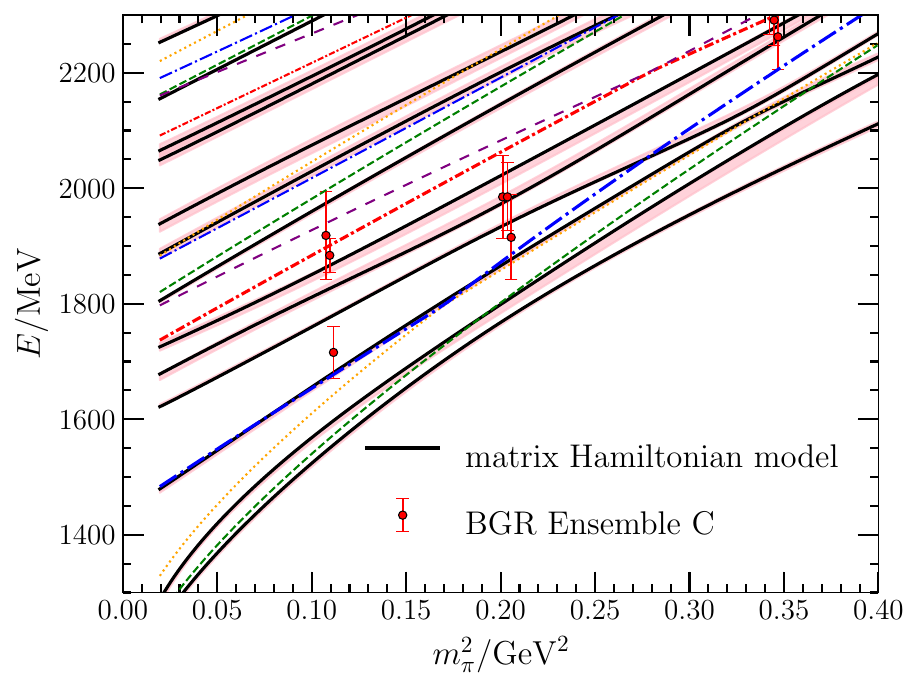}\\
\caption{Finite-volume energy spectrum with the bare baryon $\Sigma_0$ compared to the lattice QCD data with Ensemble C ($L = 2.237\,\text{fm} $) from the BGR group \cite{Engel:2013ig}. The pink shaded area illustrates the uncertainty of the HEFT results obtained by varying the Hamiltonian parameters within the allowed range as described in Sec.~\ref{jia}. See Fig.~\ref{withoutbaresp} for other captions.}
\label{cben}
\end{figure}

The BGR Collaboration also provided the lattice QCD simulations with multiple ensembles with different box sizes and quark masses, and Fig.~\ref{bben} (\ref{cben}) presents the comparison between their lattice QCD data of Ensembles B (C) and our HEFT spectrum with the inclusion of the bare $\Sigma_0$. One notices that the lattice QCD masses in Ensemble B are much more separated when the pion mass is fixed than those in Ensemble A, which indicates different states are observed in different ensembles on the lattice. Lattice QCD results were simulated on three different pion masses in Ensemble C. All these lattice QCD data are consistent with the HEFT spectra.

As shown in Fig. 5 of Ref.~\cite{Liu:2023xvy}, all the first five lowest-lying HEFT states of $\Lambda\left(\frac{1}{2}^-\right)$ were observed with much smaller errorbars by the BaSc Collaboration on the lattice \cite{BaryonScatteringBaSc:2023ori}. It will help to further constrain phenomenological models for $\Sigma\left(\frac{1}{2}^-\right)$ if the errorbars can be reduced or more states can be observed on the lattice like the $\Lambda\left(\frac{1}{2}^-\right)$ case.

\subsection{Physical $\Sigma\left(\frac{1}{2}^-\right)$ resonance spectrum}
\label{spectrum}

\renewcommand\tabcolsep{0.1cm}
		\renewcommand{\arraystretch}{1.8}
		\begin{table}[!tbp]
\caption{The pole positions of the $\Sigma\left(\frac{1}{2}^-\right)$ resonances in this work and Refs. \cite{Sarantsev:2019xxm,Zhang:2013sva,Kamano:2015hxa}. The Riemann sheets are labeled with ``u''/``p'' referred to the unphysical/physical corresponding to the channel sequence ($\pi\Lambda$, $\pi\Sigma$, $\bar{K}N$, $\eta\Sigma$, $K\Xi$). Error estimation for the scenario with bare baryon $\Sigma_0$, obtained by varying the regulator parameter $\Lambda$ within its allowed range as discussed in Sec.~\ref{jia}.
}
\centering
\label{pole}
\begin{tabular}{lllll}
	\toprule[0.3pt]
	\hline
	\hline
	  & Sheet & Pole 1 (MeV) &  Pole 2 (MeV) \\
	\hline
        Without bare baryon $\Sigma_0$ & (uuupp) & $1666-63\,i$ & - \\
        Without bare baryon $\Sigma_0$ & (uuppp) & $1637-100\,i$ & - \\
        With bare baryon $\Sigma_0$ & (uuupp) & $1687^{+0}_{-12}-110^{+3}_{+5}\,i$ & $1714^{+8}_{+1}-14^{-3}_{+6}\,i$ \\
        With bare baryon $\Sigma_0$ & (uuppp) & $1580^{+2}_{+13}-70^{+5}_{+8}\,i$ & $1707^{+6}_{+5}-17^{-8}_{+4}\,i$ \\
        Ref.~\cite{Kamano:2015hxa} Model A & - &   $1704 - 43\,i $ & -\\
        Ref.~\cite{Kamano:2015hxa} Model B & - &   $1551-178\,i$ & -\\
        Ref.~\cite{Sarantsev:2019xxm} & -  &$1689 - 103\,i$ &$1680  - 19\,i$ \\
        Ref.~\cite{Zhang:2013sva} & - &$1501 - 86 \,i$  &$1708 - 79 \,i$ \\

	\hline
	\hline
	\bottomrule[0.3pt]
\end{tabular}
\end{table}

We study the physical poles of the $\Sigma\left(\frac{1}{2}^-\right)$ resonances on the physical pion mass in this subsection. By searching for poles of the T matrices on different Riemann sheets with the parameters listed in Table~\ref{coupling}, we show the pole positions in Table~\ref{pole}. The notations ``p'' and ``u'' refer to the physical and unphysical sheets, respectively, and the sequences ``uuupp''/``uuppp'' correspond to the combinations: $\pi\Lambda$(u), $\pi\Sigma$(u), $\bar{K}N$(u/p), $\eta\Sigma$(p), and $K\Xi$(p). There is only one pole for the case without the bare baryon $\Sigma_0$, but two poles appear if the $\Sigma_0$ is included. We also list the poles from Refs.~\cite{Sarantsev:2019xxm,Zhang:2013sva,Kamano:2015hxa}. The K matrix or other parameterizations were used in Refs. \cite{Zhang:2013sva,Kamano:2015hxa} while the dynamical coupled channels framework was utilized in Ref.~\cite{Sarantsev:2019xxm} similar to this work. We can see that the poles on different sheets or with different models are not the same.

\begin{figure}[tbp]
\center
\includegraphics[width=3.4in]{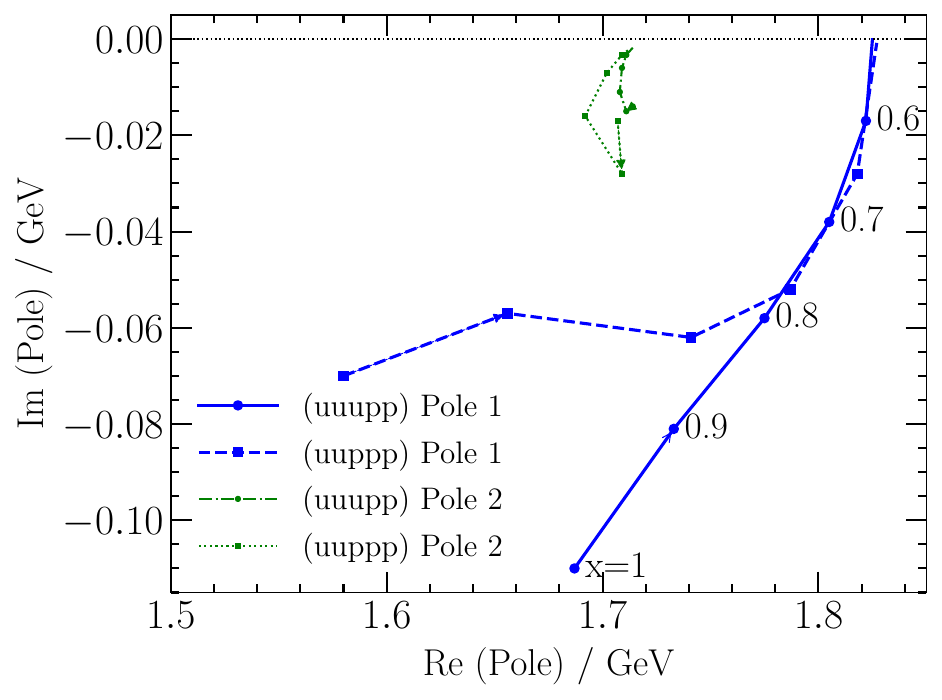}\\
\caption{The trajectories of poles in the scenario with the bare $\Sigma_0$ on different Riemann sheets when the coupling scale factor $x$ decreases. }
\label{gj}
\end{figure}

Following our earlier analysis, we focus primarily on the poles with the bare baryon $\Sigma_0$ included. The real parts of these poles are larger than the threshold of $\bar{K}N$ but smaller than that of $\eta\Sigma$, and thus the poles $1687-110\,i$ MeV and $1714-14\,i$ MeV on the ``uuupp'' sheet would determine the main behaviors of the resonances while the poles on the ``uuppp'' sheet would affect less. The latter should be the shadow poles of the former, which can be shown in Fig.~\ref{gj}. We introduce a common scaling factor $x$ to all coupling constants in Table \ref{coupling}. By gradually decreasing $x$ from 1 to 0.5, we track the pole trajectories on these two sheets. As the couplings become weaker, the imaginary parts decrease and the corresponding poles approach closer as shown in Fig.~\ref{gj}.

\begin{figure}[tbp]
\center
\includegraphics[width=3.4in]{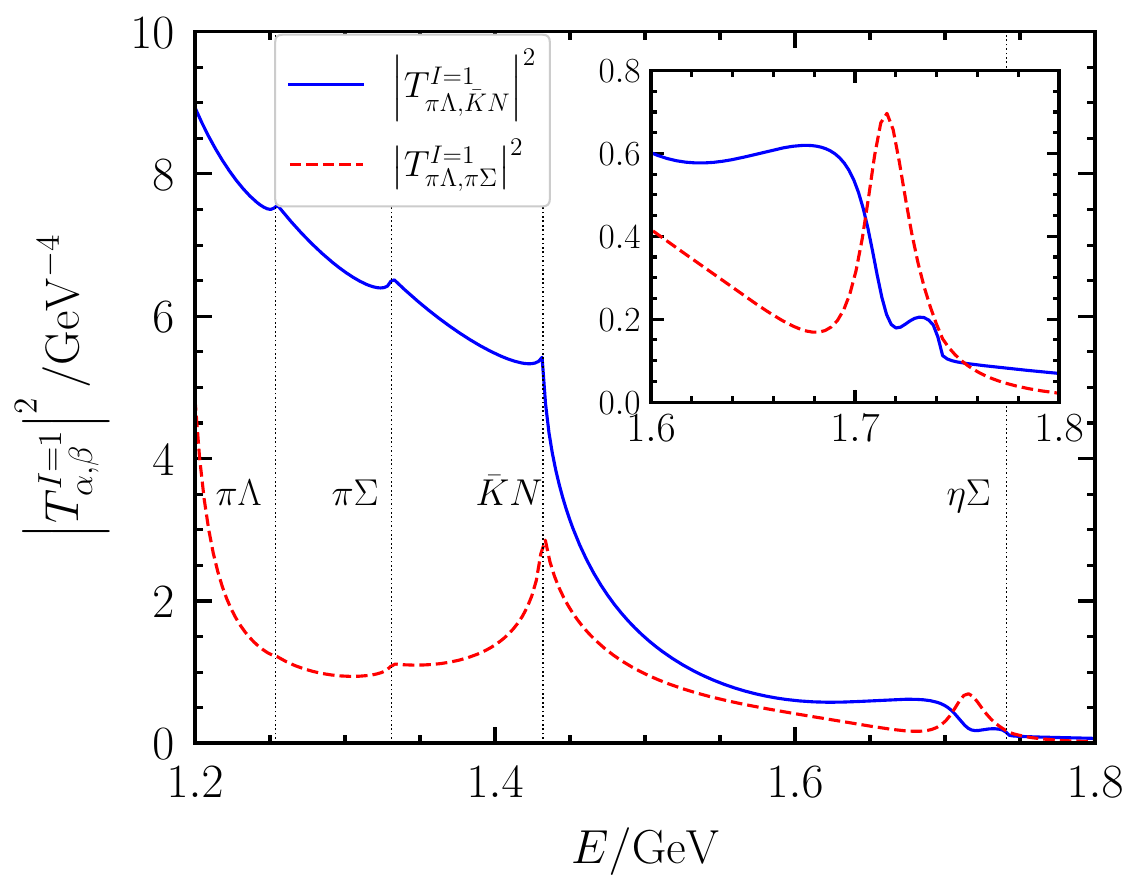}\\
\caption{The squared modules of the transition amplitudes versus the center-of-mass energy, taking  $\left| T_{\pi \Lambda, \bar{K} N}^{I=1} \right|^2$ and $\left| T_{\pi \Lambda,\pi \Sigma} ^{I=1}\right|^2$ as examples. We also insert a localized magnification of them at large energies. The vertical dotted lines indicate the thresholds of $\pi\Lambda$, $\pi\Sigma$, $\bar{K}N$ and $\eta\Sigma$.}
\label{tt}
\end{figure}

Furthermore, the Belle Collaboration reported a peak structure near the $\bar{K} N$ threshold~\cite{Belle:2022ywa}. However, due to limited experimental statistics, it is not possible to definitively distinguish whether this peak structure is caused by a resonance or a threshold cusp effect. We plot the square of T matrices versus the center-of-mass energy in Fig.~\ref{tt}, taking $\left| T_{\pi \Lambda , \bar{K} N} ^{I=1}\right|^2$ and $\left| T_{\pi \Lambda , \pi \Sigma}^{I=1} \right|^2$ as examples. Our results also show clear cusp-like structures at the $\bar{K} N$ threshold in the $\pi \Sigma\to\pi \Lambda $ scattering. Moreover, we provide a local magnification for the energy around 1.6$\sim$1.8 GeV in the subplot of Fig.~\ref{tt}. We observe a wide peak around 1.68 GeV and a narrow one around 1.73 GeV for $\left| T_{\pi \Lambda , \bar{K} N}^{I=1} \right|^2$, and there is a wide valley around 1.68 GeV and an obvious peak around 1.72 GeV  for $\left| T_{\pi \Lambda, \pi \Sigma}^{I=1} \right|^2$. These features state the presence of the two $\Sigma\left(\frac{1}{2}^-\right)$ resonances and some cusp structures. To distinguish between the cusps and the resonances, one can easily get the answer from the theoretical curves by whether the lines are smooth or not at the extreme point, but it will need much more effort in experiments. 

\subsection{SU(3) flavor symmetry and uncertainties}\label{jia}

\begin{figure}[tb]
	\includegraphics[width=3.4in]{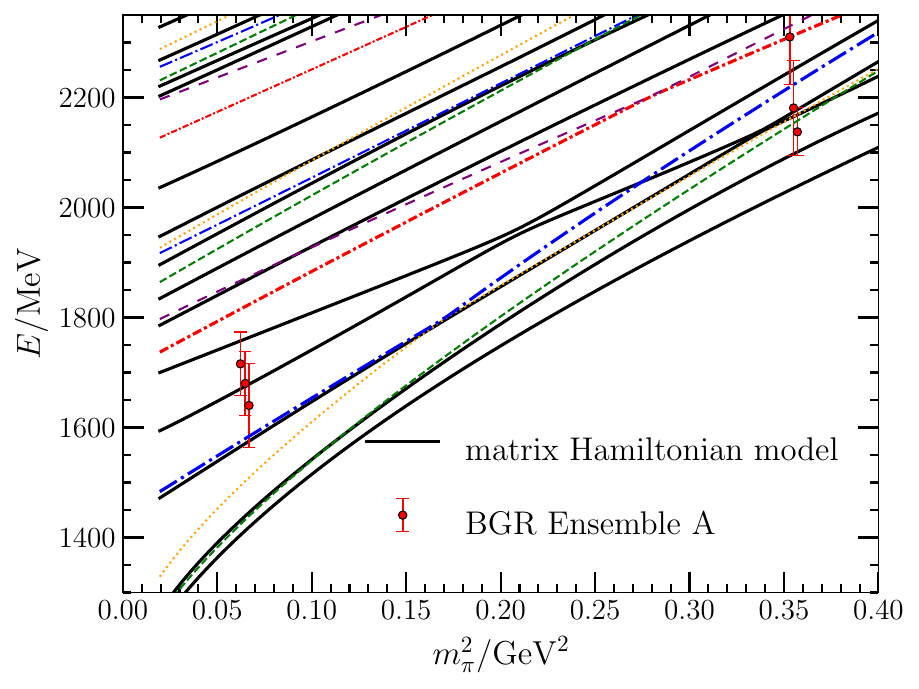}\\
\caption{Finite-volume energy spectrum  with the bare baryon $\Sigma_0$ compared to the lattice QCD data with Ensemble A ($L = 2.118\,\text{fm} $) from the BGR group \cite{Engel:2013ig}. The signs of HEFT couplings $g_{\alpha\beta}^I$ are constrained by the SU(3) symmetry. See Fig.~\ref{withoutbaresp} for other captions.
}
\label{fuhao}
\end{figure}

The Weinberg-Tomozawa potentials respect the SU(3) symmetry and can provide the couplings $g^I_{\alpha,\beta}$ \cite{Oset:1997it}. Due to the constraints of SU(3) flavor symmetry, some couplings are zero. For instance, $g_{\bar{K}N, K\Xi}^I$  are zero for both $I = 0$  and $I = 1$, and we have adopted these zero values in Ref.\cite{Liu:2023xvy}  and this work. However, since the SU(3) symmetry is obviously broken we cannot find the proper fits if strictly keeping the SU(3) symmetry. We maintain the signs consistent with the SU(3) symmetry in our previous work \cite{Liu:2023xvy}, and now we try to examine this similarity for the case of $I = 1$ and obtain the adjusted couplings $g_{\alpha, \beta}^1=\{ 0.043, 0.021, -0.061, -0.011, 0.011, -0.425, 1.000, 1.001, 0.0\\01, -0.001, -2.309, 0.001, 3.053, -0.442\}$ and bare mass $m_{\Sigma_0}=1.709$ GeV. With this set of parameters we find the poles on uuupp sheet are $1679 - 93\,i$  and $1711 - 11\,i$ MeV, which do not change very much compared to case with parameters in Table \ref{coupling}. The lattice QCD data can also be explained well as can be shown in Fig. \ref{fuhao} as the example.

The two close poles on the same Riemann sheet are very interesting and important in hadron physics. The well known two-pole structure of $\Lambda(1405)$ can be dynamically generated whether the bare $\Lambda_0$ core is included \cite{Liu:2023xvy,Liu:2016wxq} or not \cite{Ikeda:2012au,Guo:2012vv,Mai:2014xna}. Even though included, the bare core does not play an important role in forming the two poles of $\Lambda(1405)$ \cite{Liu:2023xvy,Liu:2016wxq,Hall:2014uca}. If without the $\Sigma_0$, only one pole in (1.5,1.7) GeV was found for the $\Sigma(\frac12^-)$ family in Refs. \cite{Kamano:2014zba,Kamano:2015hxa,Liu:2023xvy} and this work. The bare core is important for the two $\Sigma(\frac12^-)$ poles near 1.7 GeV. However, based on the constraints of current data, the components of these two poles still have large uncertainties. In (1.5,1.7) GeV, the lower eigenstate is mainly 65\% $K\Xi$, 25\% $\pi\Sigma$, and 10\% bare state and the upper one is 50\% $\eta\Sigma$, 23\% bare state, 18\% $\bar{K} N$, and others for the physical pion mass with $L = 2.118\,\text{fm} $ and the parameters in Table \ref{coupling}. However, if the sign constraints from SU(3) symmetry are used, the lower state becomes mainly the mixing of $\eta\Sigma$ and $K\Xi$ and the upper one is dominated by the bare state.

We use a dipole regulator to regularize the momentum integrals, and the relevant couplings are scheme dependent and would be running as the cutoff $\Lambda$ varies to keep the physical observables unchanged in principle. We allow the $\pm 0.1\,\text{GeV}$ variation for $\Lambda$ and refit the cross sections in experiments, and the corresponding changes for the parameters are shown in Table \ref{coupling}. The poles on the uuupp sheet are $1687^{+0}_{-12}-110^{+3}_{+5}\,i$ and $1714^{+8}_{+1}-14^{-3}_{+6}\,i$ MeV, and one can see that the effect from the 100 MeV change of cutoff can be most compensated by the corresponding running of parameters. For the finite volume spectra, we also add the pink shades in Figs. \ref{withoutbaresp} (b), \ref{bben} and \ref{cben}  to reflect this cutoff dependence. The error bands are obtained with $\Lambda = 1.1$ GeV  and $\Lambda = 0.9$ GeV. As can be seen from the figures, the error bands of the energy levels are narrower compared to the errors from lattice QCD simulations.

The HEFT framework establishes the correspondence between the infinite-volume scattering data and finite-volume spectra. The parameters of HEFT are determined by fitting experimental scattering data, and the success for the finite-volume spectra states that the experimental data driven procedure is powerful. We adopt the finite-range regularization which contains form factors with cutoffs, and this partly includes the effects of hadron sizes. Since the electromagnetic form factors of nucleons can be explained with the dominance of vector mesons ~\cite{Kubis:2000zd,Kubis:2000aa,HillerBlin:2017syu,Wang:2007iw,Perdrisat:2006hj,Mergell:1995bf}, the form factors in this work may contain the effective contributions of resonances we do not explicitly include. These enable HEFT to remain consistent with lattice QCD even at heavier unphysical pion masses.

\section{SUMMARY}
\label{SUM}

In this study, we utilize HEFT to discuss the $\Sigma\left(\frac{1}{2}^-\right)$ resonances below 1.8 GeV. For the internal structure of $\Sigma\left(\frac{1}{2}^-\right)$, we propose two scenarios. The first scenario postulates that the resonance contains a bare core analogous to a three-quark state in naive quark model, which mixes with the $I=1$ meson-baryon channels $\pi\Lambda$, $\pi\Sigma$, $\bar{K}N$, $\eta\Sigma$, and $K\Xi$. The second hypothesis suggests that the resonance is entirely dynamically generated by the $I=1$ meson-baryon channels without the inclusion of a bare baryon basis state. Our analysis indicates that the two scenarios are similar in describing the experimental data of low-energy $K^-p$ infinite-volume scattering cross sections. The HEFT energy levels can reproduce the lattice QCD data satisfactorily when the bare baryon $\Sigma_0$ are included,  but the HEFT fails to explain them at small pion masses in the absence of bare baryon $\Sigma_0$.

In the physical world, there are two $\Sigma\left(\frac{1}{2}^-\right)$ resonances with the masses around 1.7 GeV in our approach, which is consistent with that two lattice QCD results were obtained between the $\bar K N$ and $\eta \Sigma$ thresholds but far away from them at the small pion mass with ensemble A by the BGR Collaboration \cite{Engel:2013ig}. If the partial wave analysis improves greatly with more precise experimental measurements in the future, they can be better confirmed as well as some cusp-like structures. The lattice QCD simulations with the momentum projected two-particle meson-baryon interpolating operators and others would provide the finer spectra of $\Sigma\left(\frac{1}{2}^-\right)$ to help us further understand the physics in this strange family greatly.

\section{Acknowledgment}
This work is supported by the National Natural Science Foundation of China under Grants No. 12175091, No. 12335001, No. 12247101, the ‘111 Center’ under Grant No. B20063, and the innovation project for young science and technology talents of Lanzhou city under Grant No. 2023-QN-107.

\end{document}